\documentclass[10pt,journal,compsoc]{IEEEtran}

\ifCLASSOPTIONcompsoc
  \usepackage[nocompress]{cite}
\else
  \usepackage{cite}
\fi

\ifCLASSINFOpdf
 \usepackage[pdftex]{graphicx}
 \DeclareGraphicsExtensions{.pdf,.jpeg,.png}
\else
\fi

\usepackage{subcaption}

\usepackage{amsmath}

\usepackage{url}

\begin{document}

\title{ReviewerNet: Visualizing Citation and Authorship Relations for Finding Reviewers}
%
%
%
%

\author{Mario Salinas,
        Daniela Giorgi,
        and~Paolo Cignoni
\IEEEcompsocitemizethanks{
\IEEEcompsocthanksitem M. Salinas is with University of Pisa, 
\IEEEcompsocthanksitem D. Giorgi and P. Cignoni are with ISTI-CNR  \protect\\ 
}
}
\IEEEtitleabstractindextext{%
\begin{abstract}
We propose ReviewerNet, an online, interactive visualization system aimed to improve the reviewer selection process in the academic domain. Given a paper submitted for publication, we assume that good candidate reviewers can be chosen among the authors of a small set of relevant and pertinent papers; {ReviewerNet} supports the construction of such set of papers, by visualizing and exploring a literature citation network. Then, the system helps to select reviewers that are both well distributed in the scientific community and that do not have any conflict-of-interest, by visualising the careers and co-authorship relations of candidate reviewers. The system is publicly available, and it has been evaluated by a set of experienced researchers in the field of Computer Graphics. 
\end{abstract}

\begin{IEEEkeywords}
Scholarly data visualization, bibliometric networks, expert finding.
\end{IEEEkeywords}}

\maketitle

\IEEEdisplaynontitleabstractindextext

%
\IEEEpeerreviewmaketitle

\IEEEraisesectionheading{\section{Introduction}\label{sec:introduction}}
\IEEEPARstart{T}{he} number of digital academic documents, either newly published papers or documents resulting from digitization efforts, grows at a very fast pace: the Scopus digital repository counts more than 70 million documents and 16 million author profiles~\cite{scopus}; the Web of Science platform has more than 155 million records from over 34,000 journals~\cite{WoS}; Microsoft Academic collects about 210 million publications~\cite{MA}. In 2018, over four thousand new records were added to DBLP~\cite{DBLPrate}, and bibliometric analysts estimated a doubling of global scientific output roughly every nine years~\cite{BoMu15}. Therefore, the volume, variety and velocity of scholarly documents generated satisfies the big data definition, so that we can now talk of \emph{big scholarly data} \cite{KhLi17}. 

Sensemaking in this huge reservoir of data calls for platforms adding an element of automation to standard procedures -- such as literature search, expert finding, or collaborators discovery -- to reduce the time and effort spent by scholars and researchers. In particular, there has been an increase in the number of visual approaches supporting the analysis of scholarly data. Visualization techniques were proposed to help stakeholders to get a general understanding of sets of documents, to navigate them, and to find patterns in publications and citations. Federico et al.~\cite{FeHe17} survey about 109 visual approaches for analysing scientific literature and patents published in-between 1991 and 2016. Most of the works focused on the the visualization of document collections and citation networks. A more ambitious goal for visualization platforms would be to enable users get enough understanding to make decisions. 

In this paper, we focus on the problem of reviewer finding by journal editors or International Program Committee (IPC) members, who are required to search for reviewers who know well a subject, yet are not conflicted with the authors of the paper under scrutiny. Finding good candidate reviewers requires to analyse topic coverage (possibly during time), stage of career, and past and ongoing collaborations. Every member of the community has its own approach to reviewer finding, which usually involves bibliographic research, and frequent visits to public repositories like DBLP~\cite{ley2002dblp} and researchers' home pages. In any case, one has to confront possibly large collections of data to make decisions, and a user may easily get lost after following a few links.  

We propose ReviewerNet, a visualization platform which facilitates the selection of reviewers. The intuition behind ReviewerNet is that the authors of relevant papers are good candidate reviewers. ReviewerNet offers an interactive visualization of multiple, coordinated views about papers and researchers that help assessing the expertise and conflict of interest of candidate reviewers.

\subsection{ReviewerNet in a nutshell}

\begin{figure*}[t]
\centering
\includegraphics[width=\textwidth]{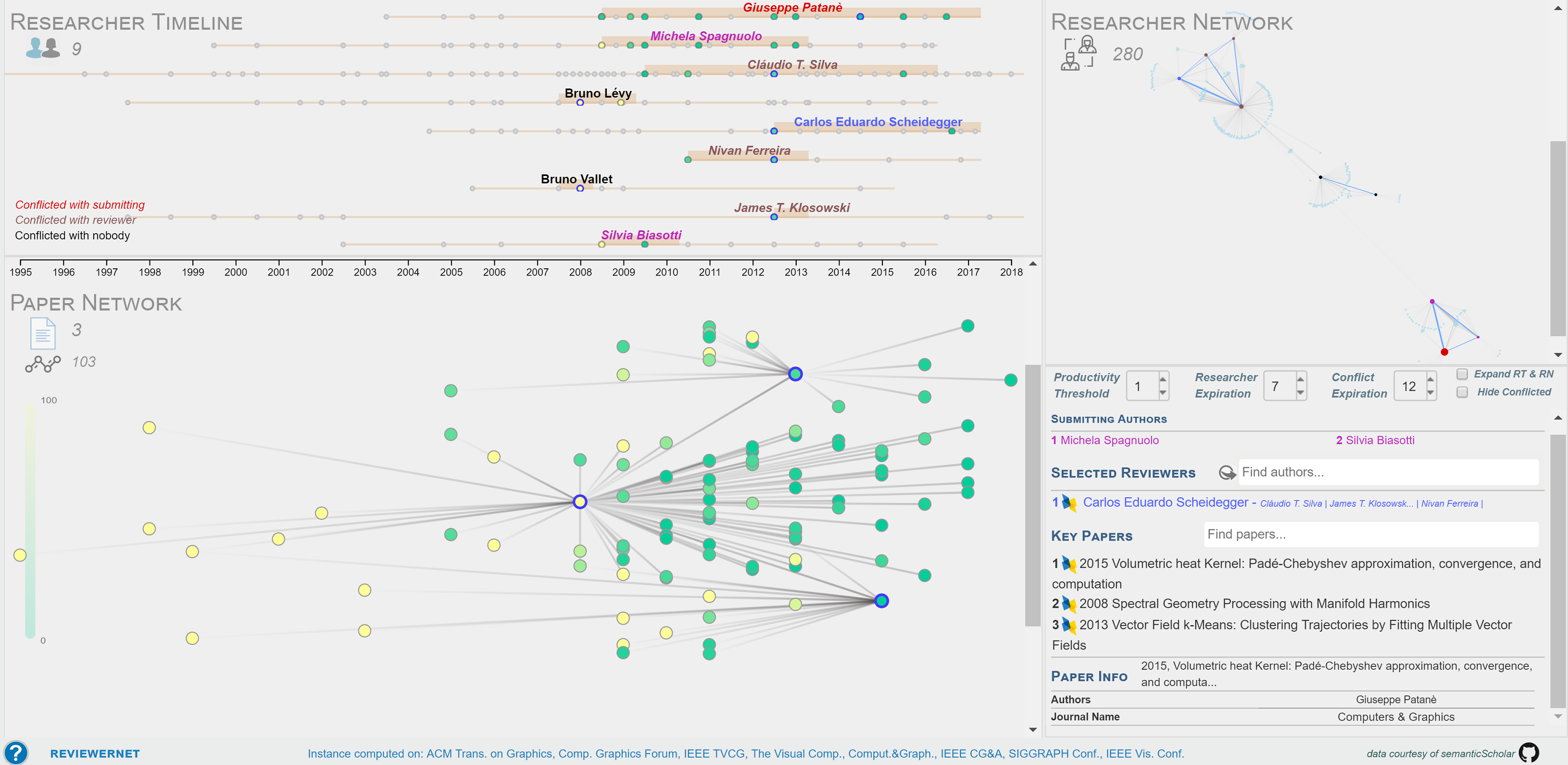}
\caption{The main interface of ReviewerNet, divided into four main areas: the \textsc{Researcher Timeline } (top left); the \textsc{Paper Network }(bottom left); the \textsc{Researcher Network } (top right); the \textsc{Control Panel } (bottom right). The interaction with these areas allows the users to identify researchers working on the topic defined by a network of papers, to analyse the researchers' contributions through time, and to get aware of co-authorship relations and conflicts.}
\label{fig:interface}
\end{figure*}

ReviewerNet supports the various actions that journal editors and IPC members perform while choosing reviewers, namely, searching the literature about the submission topic, looking for active experts in the field, and checking their conflict of interest. ReviewerNet does so by integrating an overview visualization of the literature with a visualization of the career of potential reviewers, their conflict of interests, and their nets of collaborators. This combined visualization helps to make sense of scholarly data, and rapidly get enough understanding to make a sensible decision, as shown in our user study (see Section~\ref{sec:evaluation}). 

ReviewerNet integrates the visualization of three main classes of data in a single window (see Figure~\ref{fig:interface}):

\begin{itemize}
\item  {\bf Paper Network} (PN): a chronologically ordered visualization of the literature citation network related with the submission topic. The nodes represent papers, while arcs represent in- and out-citation relations between papers. The horizontal dimension represents time. By means of interactive graph expansion functionalities, the PN supports the rapid exploration of key papers in the literature with respect to the topic of the submitted paper. The authors of the key papers identified will define the set of the candidate reviewers. The PN is built by the users, starting from a small number of seed papers of their choice;
\item  {\bf Researcher Timeline} (RT): a time-based visualization of the academic career of researchers, through horizontal lines and bars. The RT helps assessing the suitability of potential reviewers, showing thier topic coverage, productivity over years, and stage of career. Also, visual cues help the user to tell apart candidate reviewers from conflicting researchers. The RT is built automatically by ReviewerNet while the user builds the PN;
\item  {\bf Researcher Network} (RN), a graph visualization of co-authorship relations: the nodes represent the authors in the PN and their collaborators in the dataset; the arcs connect authors who have publications in common. The aim of the RN is to visualize the research communities: indeed, the identification of network of collaborators helps looking for sets of independent, non-conflicting reviewers. As with the RT, the RN is built online by ReviewerNet.
\end{itemize}
The basic pipeline for finding reviewers with ReviewerNet involves building the Paper Network starting from a small set of seed documents; evaluating possible choices of reviewers, by navigating the Reviewer Timeline and the Reviewer Network; and finally obtaining a justified list of chosen reviewers, along with possible substitutes suggested by ReviewerNet in case of decline.   

The user can navigate the different views and interact with the system through simple actions, to drive his/her investigation. Each view in ReviewerNet is linked to the other views, so that so that any action in a view is reflected in the others. Visual cues are used to improve the comprehension during interactive sessions: the colour, size, boundary, and style of visual elements visually represent important characteristics of the entities they stand for. Moreover, the coherence of visual cues across different views enforces their meaningfulness, and makes it easy for the user to switch between different views without losing focus.  

ReviewerNet builds on a reference database including papers, authors and citations from selected sources (journal articles and conference papers) taken from the Semantic Scholar Research Corpus~\cite{ammar:18}. ReviewerNet can be built over any dataset, according to the domain of interest.

\subsection{Summary of contribution}

We introduce the ReviewerNet visualization platform, which supports the reviewer selection process in the academic domain (Section~\ref{sec:methods} and Section~\ref{sec:userinterface}). 

We demonstrate the platform in the field of Computer Graphics, with a reference dataset containing 17.754 papers, 108.155 citations, 23386 authors. We show how ReviewerNet can be used to search for reviewers who are expert on a certain topic, are at a certain career stage, who have a certain track of publishing records, who are not conflicting with neither the submitters nor other reviewers, and who are well-distributed in the scientific community (Section~\ref{sec:demonstration}). 

We evaluate the platform through a user study involving 15 real end-users from the Computer Graphics community, and show how they were able to get acquainted with ReviewerNet even with a very limited training, and how they rated very positively ReviewerNet functionalities (Section~\ref{sec:evaluation}).  

One of the main advantages of ReviewerNet is that it only relies on citations, to analyse the literature, and on co-authorship relations, to analyse conflicts. Citations are an essential part of research: they represent a credible source of information about topic similarity and intellectual influence. Moreover, since citations have author-chosen reliability, they are a very robust cue to relatedness. Similar reasonings hold for co-authorship relations. Therefore, an important contribution is the demonstration that a well-combined visualization based on citation and co-authorship relations only can support the reviewer search process, without the need for more complicated content analysis techniques.  

The tool is free to be used and open source; the source code is available at~\url{https://github.com/cnr-isti-vclab/ReviewerNet}.

\section{Related work}
\label{sec:related}

Concerning the reviewer selection process, the literature mostly focused on the automatic reviewer \emph{assignment} task, which is a different problem than ours. Indeed, the reviewer assignment problem requires finding the best assignment between a finite set of reviewers (e.g., the members of the Programme Committee of a conference) and a finite set of papers (the papers submitted to the conference); this is usually done using bi-partite graph matching and taking into account pertinence of the reviewers with the papers and fair distribution of loads; \cite{WaCh10} provides an overview of this problem. 

In what follows, we briefly review the state-of-the art about the search, analysis and recommendation services offered by scholarly data platforms (Section \ref{sec:schoplat}), and the visualization of bibliometric networks (Section \ref{sec:bibvis}). 

\subsection{Scholarly data platforms}
\label{sec:schoplat}
Many applications have been developed on top of the big scholarly data platforms to search for authors, documents, venues, and analyse statistics about for example distribution per research area, citations, and other bibliometric indices. Most academic search engines also provide research paper recommendations according to one's research interests. 

Microsoft Academic provides a semantic search engine that employs natural language processing and semantic inference to retrieve the documents of interest. It also provides related information about the most relevant authors, institutions, and research areas \cite{SiZh15}. Scopus enables one to search for authors or documents,  track citations over time for authors or documents, view statistics about an author's publishing output, and compare journals according to different bibliometric indices \cite{scopus}. 

These and similar applications offer basic functionalities and static visualizations which researchers do use while looking for reviewers. Though, none of them offers an integrated service to support the higher level tasks of fine-tuned reviewer selection, where both expertise and conflicts of interest have to be taken into account. 

\subsection{Visualization of bibliometric networks}
\label{sec:bibvis}
The visualization of bibliometric networks is an active area of research \cite{Ch13,FeHe17}. Bibliometric networks include citation, co-citation, co-authorship, bibliographic coupling and keyword co-occurrence networks. 

Concerning visualization of citations, most part of the literature focused on co-citation and bibliographic coupling networks, rather than on direct citations. One of the first visualization of citation networks is Garfield's historiography \cite{GaPu03}, a node-link diagram where citation links are directed backwards in time. Garfield and colleagues underline how citation networks enable one to analyse the history and development of research fields. CiteNetExplorer \cite{vEWa14} is a software tool to visualize citation networks which builds on Garfield and colleagues' work: it improves the graph layout optimization to handle a larger number of papers, and offers network drill-down and expansion functionalities. PaperVis \cite{ChYa11} is an exploration tool for literature review, which adopts modified Radial Space Filling and Bullseye View techniques to arrange papers as a node-link graph while saving the screen space, and categorizes papers into semantically meaningful hierarchies. 

\cite{GoLi13} describes a visual analytics system for exploring and understanding document collections, based on computational text analysis; it supports document summarization, similarity, clustering and sentiment analysis, and offers recommendations on related entities for further examination. Rexplore \cite{OsMo13} is a web-based system for search and faceted browsing of publication. Rexplore also includes a graph connecting similar authors, where similarity depends on research topics as extracted from document text. At any rate, using keywords as proxies for research topics can be noisy. Therefore, in ReviewerNet we only rely on co-authorship relations.    

Many of the approaches for bibliographic network visualization make limited use of user interaction, and often use a loose coupling of views \cite{FeHe17}. With ReviewerNet, we propose an integrated environment which facilitates a high-level task (reviewer discovery and selection) by means of coordinated, interactive views. Also, only a few works include an in-depth evaluation of the techniques proposed through user studies. We report a user study involving real end-users, namely 15 experts in Computer Graphics, who tested ReviewerNet and filled in an anonymous questionnaire.

\section{Objectives}\label{sec:methods} 

The aim of REviewerNet is to facilitate the reviewer selection process in the academic domain. While designing ReviewerNet, we took into account the characteristics of the \emph{users}, the users' \emph{tasks}, and the \emph{data} the users are working with \cite{MiAi14}. 

The \emph{users} of ReviewerNet are researchers, and in particular those playing the role of journal editors, associate editors, and members of IPCs of big conferences, in any field of research. Their \emph{task} is searching for reviewers for a submitted paper: this involves searching the literature for key papers and authors in the field; evaluating the candidates' research interests and their evolution over time; and assessing the candidates' conflict of interest with respect to the submitting authors and other reviewers. ReviewerNet supports all these subtasks, by visualizing the literature related with a topic, the career of relevant researchers in the field, and the relationships among researchers. 

The data pertain to three types of entities: \emph{papers}, \emph{researchers}, and \emph{citations}. The data attributes are both quantitative and qualitative, and the time dimension is central. 

In ReviewerNet, the attributes of a paper which are visualized are its \emph{citation count} -- the number of papers citing it -- as well as standard \emph{bibliographic attributes} -- title, authors, publication year, venue. Papers are related through \emph{direct citations}. 

Researchers have two attributes in ReviewerNet: relevance and conflict of interest. We define a researcher's \emph{relevance} as a reviewer according to the authorship of relevant papers. The concept of relevance can be tuned according to the user needs (e.g., looking for highly-specialized reviewers, as opposed to generalists). The second attribute of researchers is their \emph{conflict of interest}, with either the submitting authors or other reviewers. We model the conflict of interest after \emph{co-authorship} relations: two researchers have a conflict of interest if they have papers in common. We let the degree of conflict, and hence the availability as a reviewer, be modulated according to the number of papers in common, and the years passed since the last co-authored paper, again according to the user intent. 

The following section details the notation used in the rest of the paper, and the formal definition of paper and researcher attributes. 

\subsection{Notation}

Let $\mathcal{P}$ denote the set of papers in a reference dataset, 
and let $\mathcal{P}_{V} \subseteq \mathcal{P}$ be the set of papers relevant to a submission. 
$\mathcal{P}_{V}$ is built by the users starting from a small number of seed papers of their choice (cf. Section \ref{sec:demoPN}).

A paper $p \in \mathcal{P}_{V}$ is marked as \emph{selected}, if it is considered as a key paper by the user; we denote by $\mathcal{P}_{S}$ the set of selected papers, with $\mathcal{P}_{S} \subseteq \mathcal{P}_{V} \subseteq \mathcal{P}$. 

If $\mathcal{C}(p)$ is the set of papers citing $p$, the \emph{citation count} $c(p)$ is its cardinality: $c(p) = \vert \mathcal{C}(p) \vert$.

Let $\mathcal{A}(p)$ be the set of authors of a given paper $p$, and $\mathcal{R}$ the set of authors of papers in $\mathcal{P}$. 
Then, the set $\mathcal{R}_C \subseteq \mathcal{R}$ of \emph{candidate reviewers} is given by the set of researchers who authored a selected paper: $$\mathcal{R}_{C} = \{r \in \mathcal{R} \ s.t. \ \exists \ p \in \mathcal{P}_S : r \in \mathcal{A}(p)\}$$ 
For a candidate reviewer $r$, let $\mathcal{P}_{S}|_{r}$ be the set of papers in $\mathcal{P}_{S}$ authored by $r$. Then, the \emph{relevance score} $s(r)$ of the candidate reviewer $r$ is defined as a weighted sum of the number of selected and non-selected papers in $\mathcal{P}_V$ authored by $r$: $$s(r) = \alpha \vert \mathcal{P}_{S}|_{r} \vert + \beta \vert \{\mathcal{P}_{V} - \mathcal{P}_{S}\}|_{r}\vert$$ with $\alpha$ and $\beta$ real-valued coefficients summing up to one. We set $\alpha = 0.7$ and $\beta = 0.3$ as default parameters. The set of candidate reviewers will be visualized in the Researcher Timeline in order of their relevance; relevance will also define the dimension of nodes in the Researcher Network.

Finally, $\mathcal{CA}(r)$ denotes the set of co-authors of a researcher $r$, or, in other words, the set of researchers who have a conflict with him/her.

\section{User interface}
\label{sec:userinterface}

There are four regions in the user interface, described below and shown in Figure~\ref{fig:interface}. Each region is resizable in height. The visual composition helps the user to gain different perspectives on the problem at hand, within a single visualization. \\

\noindent{\bf The Paper Network:} 
The Paper Network (PN), at the bottom-left hand side of the screen, is a graph visualization of the literature relevant to a submission topic. The nodes represent papers in $\mathcal{P}_{V}$, while the arcs represent in- and out-citation relations between them. The horizontal dimension represents time, as papers are ordered according to their publication year. A force-directed graph drawing algorithm determines the layout in the vertical direction. The Paper Network is built interactively by the user. 
\\

\noindent{\bf The Researcher Timeline:} 
The Researcher Timeline (RT), at the upper-left side of the screen, is a visualization of the academic career of researchers, through lines and bars. Each line represents a candidate reviewer $r$ in $\mathcal{R}_{C}$, that is, the author of a selected paper in $\mathcal{P}_{S}$. The dots over the line represent the set $\mathcal{P}|_{r}$ of papers authored by $r$ in the reference database $\mathcal{P}$. The Researcher Timeline is constructed and updated automatically by ReviewerNet while the user builds and refines the Paper Network.  \\

\noindent{\bf The Researcher Network:} 
The Researcher Network (RN), at the upper-right hand side of the screen, is a graph visualization of the co-authorship relations. The nodes are the researchers in $\mathcal{R}_V$ along with their collaborators in $\mathcal{R}$. The arcs connect authors who have publications in common: for each node representing a researcher $r$, the node degree is the cardinality $\vert \mathcal{CA}(r) \vert$. A force-directed graph drawing algorithm determines the graph layout. As with the Researcher Timeline, the Researcher Network is built automatically by ReviewerNet while the user builds the Paper Network.  \\

\noindent{\bf The Control Panel:} 
The Control Panel (CP), at the bottom-right hand side of the screen, allows the user to input and manage the names of submitting authors, the names of selected reviewers, and the titles of key papers. The CP area also displays information about papers, upon request. The DBLP icon beside reviewers' names and paper titles links to their respective DBLP page. Moreover, the CP includes parameters boxes and checkboxes to fine-tune the visualization (cf. Section \ref{subsec:parameters}). Finally, the CP enables the user to download the list of selected reviewers, along with substitute reviewers suggested by ReviewerNet.  

\begin{figure*}[!pt]
\centering
\includegraphics[width=\textwidth]{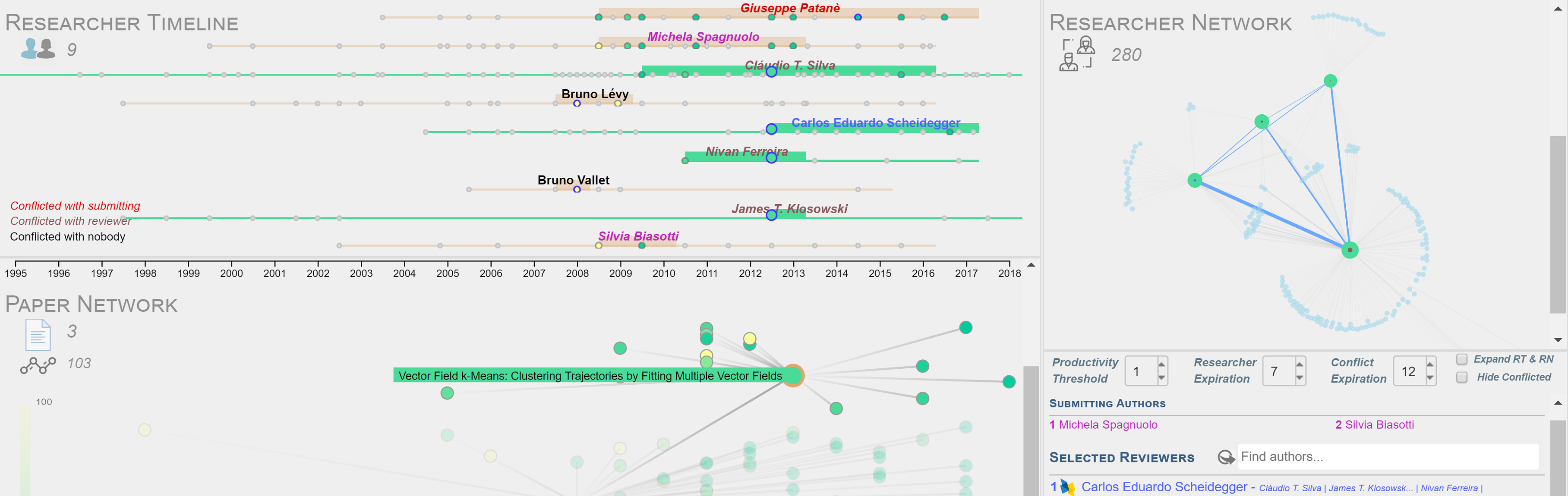}
\caption{When hovering over an entity representing a paper, the authors of that paper are highlighted in the other views.}
\label{fig:paperhovering}
\end{figure*}

\subsection{Visual consistency}
\label{subsec:visualvar}

Visual cues include the colour, size, boundary, and style of visual elements representing papers, researchers and their relations across the different views.  \\

\noindent{\bf Visual cues for papers:} 
For a paper $p \in \mathcal{P}_{V}$, the color corresponds to the citation count $c(p)$, from yellow (few citations) to green (many citations). This colormap applies to both nodes in the PN and dots in the RT. Dots corresponding to papers in $\mathcal{P} - \mathcal{P}_V$ (papers in the reference database, but not included the PN) are marked as grey. 

Selected papers in $\mathcal{P}_S$ are circled in blue, both in the PN and the RT.  Arcs are blue in the RN when the co-authored papers include a selected paper.  \\

\noindent{\bf Visual cues for researchers:} 
For researchers in the RT, the name coloring emphasizes the distinction between roles: submitting authors (marked as purple), their co-authors (red), selected reviewers (blue), their co-authors (brown), and non-conflicting, candidate reviewers (black). The nodes in the RN corresponding to researchers in the RT follow the same rule, whereas nodes representing their co-authors in $\mathcal{R}$ are light blue.   

For researchers in the RT, the font style of names further helps to tell apart conflicting researchers (italic) from non-conflicting candidate reviewers (normal). The same colour/font rules apply to the names suggested in the selected reviewers' drop-down menu in the CP.

The researchers in the RT are ordered vertically according to their relevance score $r(s)$. The same score is rendered in the RN through the dimension of nodes.  \\

\begin{figure*}
\centering
\includegraphics[width=\textwidth]{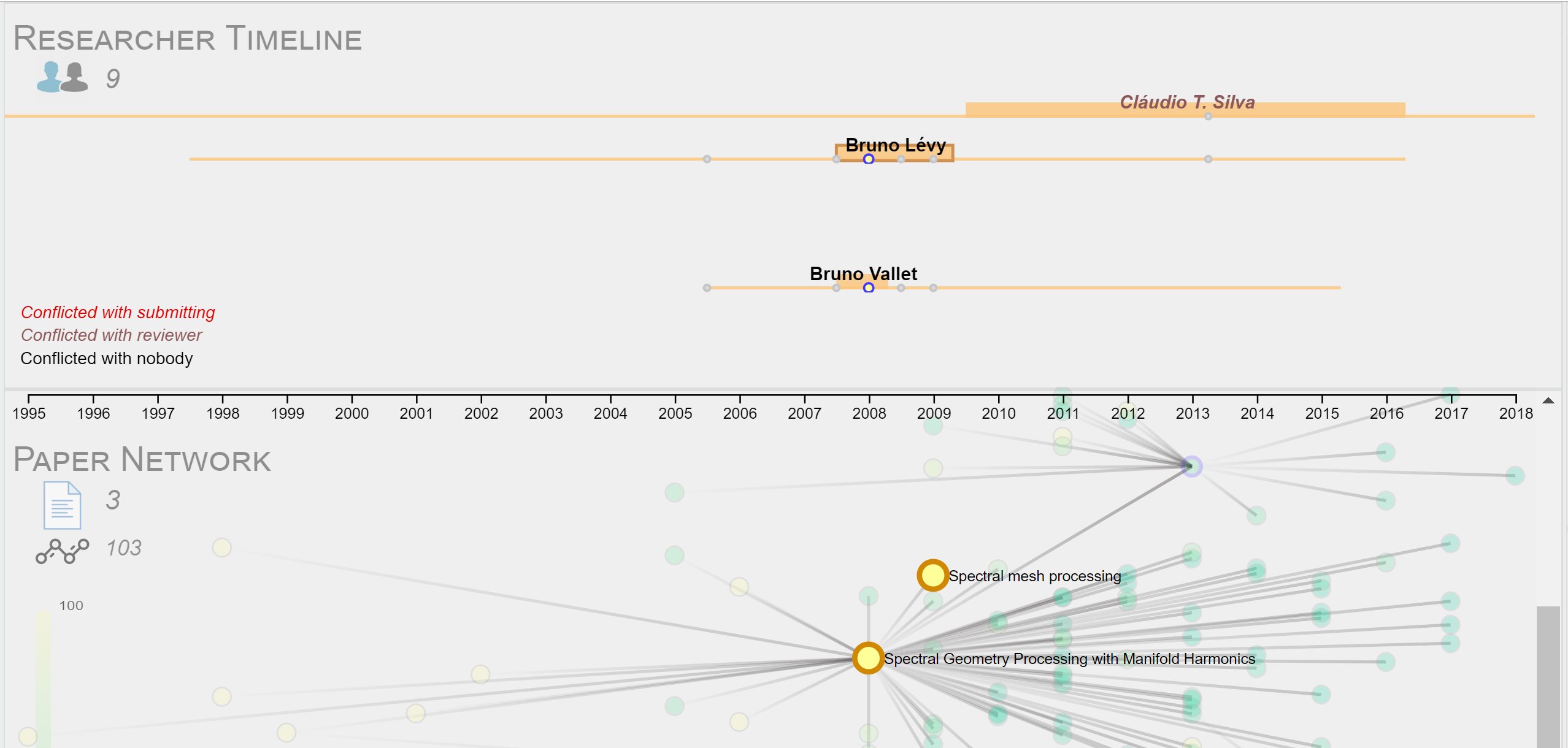}
\caption{Focusing on a researcher by clicking on her/his name in the Researcher Timeline allows to highlight her/his co-authors and production in the Paper Network.}
\label{fig:co-authors}
\end{figure*}

\begin{figure*}
\centering
\includegraphics[width=\textwidth]{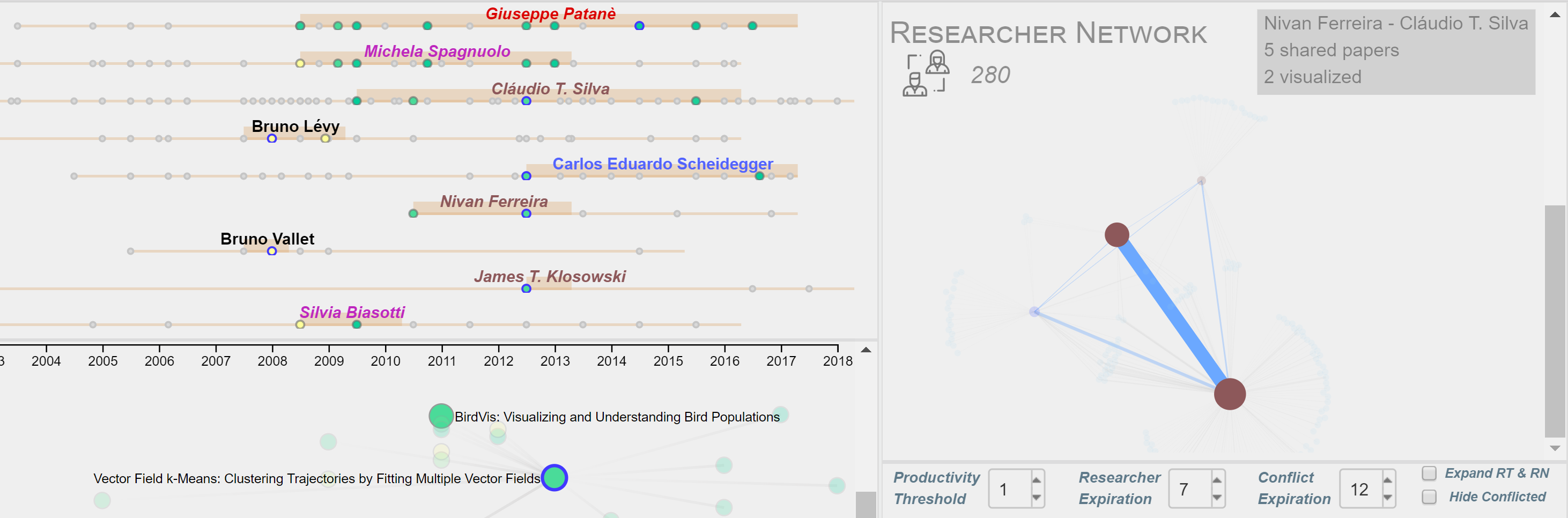}
\caption{Hovering over a segment joining two researchers in the Researcher Network shows details about their co-authored papers and highlights them in the Paper Network.}
\label{fig:researchernetwork}
\end{figure*}

\begin{figure}
\centering
\includegraphics[width=0.5\textwidth]{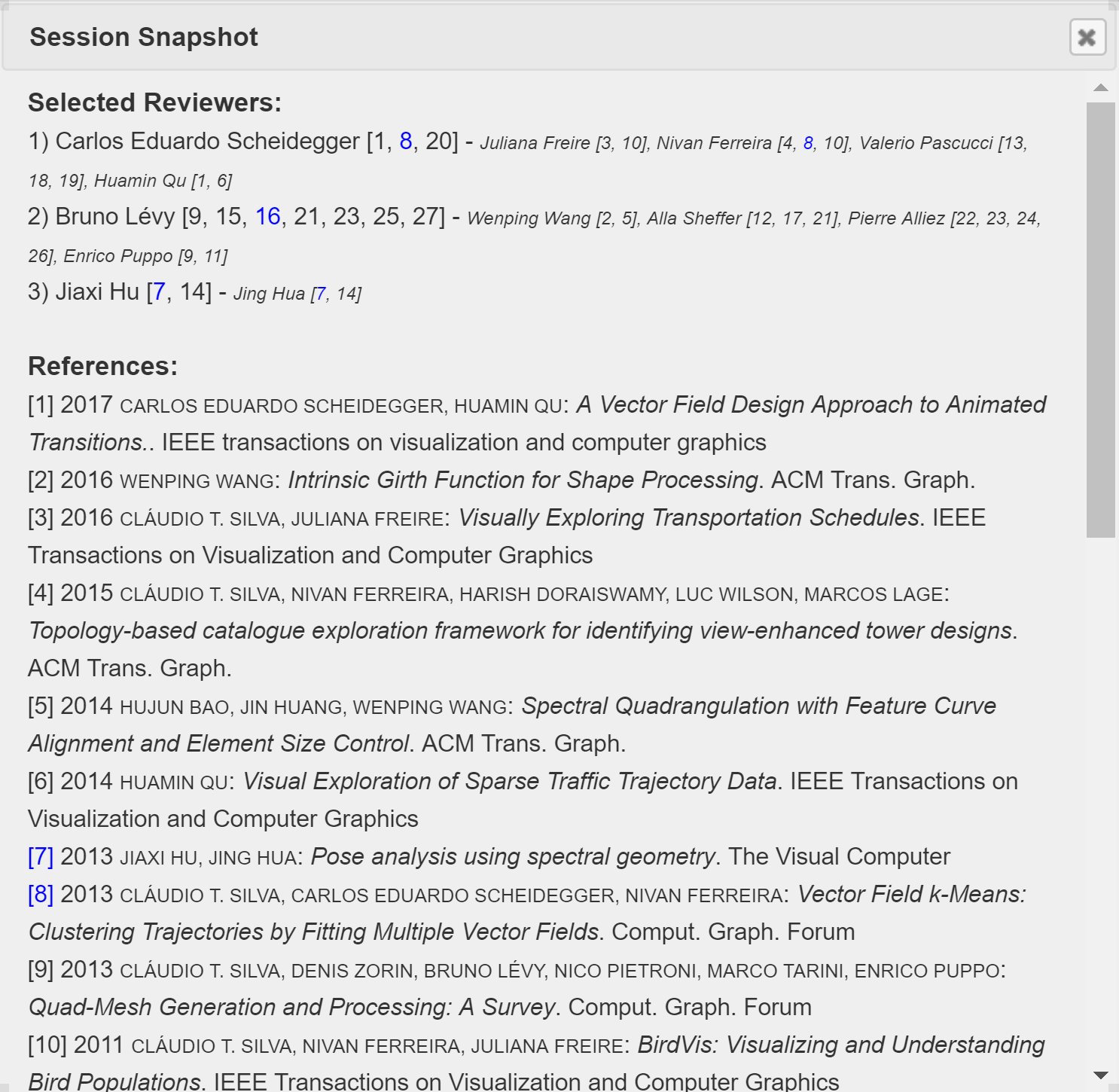}
\caption{The list of selected reviewers together with substitutes and a bibliography. For each selected reviewer we list the papers he/she has authored and that motivated his/her choice as a reviewer.  The substitute reviewers are those researchers that have authored a similar set of publications and have the same conflicts as the selected reviewer, and therefore can replace him/her in case of decline.}
\label{fig:list}
\end{figure}

\subsection{Actions}
\label{subsec:actions}

Each view (PN, RT, RN, CP) is linked to the other views, so that any action in a view is reflected in the others. \\

\noindent{\bf Actions on Papers:} 
The first step is to build the Paper Network, that is, a set of key papers which are relevant to the submission topic. The user inizializes the PN by inputing the titles of a small set of seed papers in the \emph{Key papers} field, with the help of title-based suggestions. The seed papers are visualized in the PN, along with their in- and out-citations. The user can now expand the network, to discover additional documents. With a double click, he selects interesting nodes, i.e., papers he/she deems relevant to the submission topic. The PN then updates with the in- and out-citations of the selected papers. Papers can be deselected with a double click. 

When the users focuses on a paper in one of the views by mouse hovering, the same paper is highlighted in the other views. For example, when hovering the mouse over a node in the PN, the corresponding dot in the RT is highlighted, and viceversa. Also, the paper details (title, publication year, venue) are shown in the CP on a mouse click. Likewise, by hovering over or clicking on the title in the CP, the corresponding node and dot are highlighted in the PN and the RT.
       
When hovering the mouse over an entity representing a paper (a node in the PN, a dot in the RT bars, the title in the CP), the paper authors are highlighted in the RT and RN, if present (Figure \ref{fig:paperhovering}). A mouse click on the focused paper lets the user navigate the visualization with highlighted items. A single click restores the previous visualization.

The icon beside the paper title in the CP links to the DBLP page of the paper. \\

\noindent{\bf Actions on Researchers} In a similar fashion to papers, when the user focuses on a researcher in one of the views by mouse hovering, the same researcher is highlighted in the other views. When hovering the mouse over a node in the RN, the name of the corresponding researcher appears on the upper-right corner.  

When hovering the mouse over an entity representing a researcher (a bar in the RT, a dot in the RN, the name in the CP), the papers authored by the researcher are highlighted in the PN view. 

A mouse click on a researcher puts the focus on him/her, his/her production and his/her personal net of collaborators (Figure \ref{fig:co-authors}). The user can navigate a visualization with selected items and additional functionalities. Only the set of co-authors is visualized in the RT and the RN. While hovering on one of the co-authors, the common publications are shown in the PN, and the arc representing the co-authorship relation is visualized in the RN. Another mouse click will get the user back the previous visualization.

When hovering the mouse over an arc in the RT, a pop-up on the upper-right corner shows the pair of co-authors names, the number of common papers in the dataset $\mathcal{P}$, and the number of common relevant papers in $\mathcal{P}_{V}$. In turn, for blue arcs, the common papers are highlighted in the PN (Figure \ref{fig:researchernetwork}). 

The icon beside the researcher name in any of the fields in the CP links to the DBLP page of that researcher. 

A researcher can be removed from the list of selected reviewers with a double click.

Finally, for each reviewer selected by the user, ReviewerNet suggests a set of possible substitute reviewers, in case of negative answers from the selected one. For a selected reviewer $r$, the alternative reviewers are chosen in the set $\mathcal{R}_C$ of candidate reviewers, so that they only conflict with $r$, and with no other selected reviewer; the list of substitutes is ordered according to the number of common papers between the reviewer and his/her substitute. The user can exchange a reviewer with one of his/her substitutes by clicking on the name of the substitute.

Work sessions can be saved for later re-use and re-assessment.

The export button enables the user to download the list of reviewers and their potential substitutes. The list also includes references to the reviewers' publications in the dataset $\mathcal{P}$ (Figure \ref{fig:list}).  

\subsection{User-defined parameters and settings}
\label{subsec:parameters}

Users can adjust the number of candidate reviewers visualized through a set of thresholds and options. \\

\noindent{\bf Size of data visualized:} 
To limit the number of candidate reviewers visualized in the RT and the RN, the user can set two thresholds a researcher has to meet to be considered as a candidate reviewer: 
\begin{itemize}
\item \emph{Productivity threshold}: the minimum number of authored selected papers in $\mathcal{P}_S$ (i.e., $\vert \mathcal{P}_{S}|_{r} \vert$ has to be greater than the threshold, for a researcher $r$ to be included in the set $\mathcal{R}_C$ of candidate reviewers);
\item \emph{Researcher expiration}: the maximum number of years since the last authored paper in the reference dataset $\mathcal{P}$ (i.e., the number of years has to be lower than the threshold for a researcher to be considered active and included in $\mathcal{R}_C$).
\end{itemize}
The user can also remove conflicting authors and their co-authors from the visualization, by ticking the \emph{Hide Conflicted} checkbox.
To augment instead the number of potential reviewers visualized, the user can tick the \emph{Expand RT \& RN} checkbox: the visualization will include all the researchers in $\mathcal{R}_V$ (all the authors of relevant papers) instead of the researchers in $\mathcal{R}_C$ only (the authors of selected papers only). Note that visualizing a large number of researchers can slow down the interface. \\

\noindent{\bf Conflict-of-interest:} 
Finally, to modulate the conflict of interest, the user can set a threshold for two researchers to be considered as co-authors, namely 
\begin{itemize}
\item \emph{Conflict expiration}: the maximum number of years since the last co-authored paper in $\mathcal{P}$. 
\end{itemize}
A larger threshold will increase the number of candidates marked as conflicted. Conversely, a smaller threshold will increase the number of available reviewers.

\section{Demonstration}\label{sec:demonstration}
To better explain how ReviewerNet works and supports the reviewer selection process, this section presents an example user scenario. We introduce Robert, a fictitious academic researcher. Robert is in the IPC of a conference in the field of Computer Graphics; he is the primary reviewer for a paper, and he is in charge of finding three additional reviewers, plus alternative reviewers in case of decline. 

Below we describe Robert's interaction with ReviewerNet. In addition, since a static description may not adequately convey the dynamic nature of Robert's investigation, we refer the reader to the accompanying video at \url{https://www.youtube.com/watch?v=JnomPO8QI28}, which illustrates the scenario described below. 

The demonstration platform is available at \url{https://reviewernet.org/}.  

\subsection{Data collection}

To construct the reference dataset for this scenario, we collected papers, authors and citations from eight selected sources in the field of Computer Graphics, taken from the Semantic Scholar Research Corpus \cite{ammar:18}. The dataset includes data from the journals and conference proceedings listed in Table \ref{table:sources}, spanning the years in-between 1995 and 2018. After an automatic cleaning steps to remove non-papers (such as acknowledgments to reviewers, prefaces, etc.), the final reference dataset contains 17.754 papers, 108.155 citations, and 23386 authors. 

\begin{table}[!t]
\renewcommand{\arraystretch}{1.3}
\caption{The selected sources from the Semantic Scholar Research Corpus used in our demonstration scenario. The final reference dataset contains 17.754 papers, 108.155 citations, and 23386 authors.}
\label{table:sources}
\centering
\begin{tabular}{|l|c|}
\hline
ACM Transactions on Graphics & 2833\\ 
Computer Graphics and Applications  & 1983 \\ 
Computer Graphics Forum & 3238\\ 
Computers \& Graphics & 2155\\ 
IEEE Transactions on Visualization and Computer Graphics & 3236\\ 
Visual Computer & 2107\\ 
Proceedings of IEEE Conference Visualization (pre 2006) & 501 \\ 
Proceedings of ACM SIGGRAPH (pre 2003) & 1701\\
\hline
\end{tabular}
\end{table}

\subsection{ReviewerNet in action}

Robert is in charge of finding reviewers for a paper about polycube maps, authored by Marco Tarini and Daniele Panozzo. He inputs thir names in the \emph{Submitting Authors} field (also with the help of the drop-down menu), and ticks the \emph{Done} checkbox. The authors are now shown in the Researcher Timeline and the Reviewer Network, marked as purple, and the rest of the interface becomes active. 

\subsubsection{Building the Paper Network} 
\label{sec:demoPN}
The first step is to build the Paper Network, that is, a set of key papers which are relevant to the submission topic. Later on, Robert will chose his reviewers among the authors of those key papers. Robert thinks of a first set of three documents about polycube maps, which serve as seeds for building the network (\emph{PolyCube-Maps}, 2004; \emph{A divide-and-conquer approach for automatic polycube maps construction}, 2009; \emph{$L_1$-based construction of polycube maps from complex shapes}, 2014). He inputs their titles in the \emph{Key papers} field. His knowledge of the domain helps him in this initial step, though he can also take advantage of title-based suggestions, which are shown in a drop-down menu, listed by publication year. The three papers are now included in the Paper Network, along with their in- and out-citations. 

Robert can now expand the network, to discover additional documents. With a double click, he selects interesting nodes, i.e., papers he deems relevant to polycube maps. The Paper Network then updates with the in- and out-citations of the selected papers, so that Robert can further explore the literature. Robert navigates the network, and decides to reduce its size by deselecting a paper he realizes he is no longer interested in, because its citations suggest it addresses a different topic than the submission. Selected papers are marked with a blue contour circle, both in the Paper Network and the Researcher Network. 

Robert continues until he feels the selected papers and their citations offer a good coverage of the literature about the topic at hand. Robert checks the paper details, including the link to the respective DBLP page, shown in the bottom right corner of the interface. A quick keyword search with \emph{polycube maps} in the \emph{Key papers} field let him notice that there is an important paper he was missing (\emph{Efficient volumetric poly-cube map construction}, 2016); the paper can be easily told apart from papers already in the network, thanks to visual cues in the drop-down menu. 

While Robert builds his Paper Network, ReviewerNet automatically adds the authors of selected papers in the Researcher Timeline and the Researcher Network, as candidate reviewers.  The selection of 6 papers produces a list of 28 candidate reviewers.

\subsubsection{Exploring the Researcher Timeline and the Researcher Network} 
 
Robert now explores the Researcher Timeline to assess the suitability of candidate reviewers. In the Researcher Timeline, researchers are represented as horizontal lines, spanning their academic career. Robert checks the expertise of candidate reviewers by looking at their stage of career, and production over years. Since each view is linked to the other views, Robert checks topic coverage by looking at who published what, by hovering the mouse over papers to highlight their authors in all the views. He checks conflicts with the submitting authors, thanks to colours and font style. 

The visualization also help Robert analysing the network of collaborators of candidate reviewers. This is fundamental to find sets of independent, well distributed reviewers. With a mouse click on a researcher, ReviewerNet highlights his/her co-authors, and highlights on demand the common publications. Robert further investigates on the collaborations among candidate reviewers by navigating the Researcher Network, a graph visualization of co-authorship relations among the candidate reviewers and their collaborators in the dataset. Robert pans and zooms and uses the different handlers available to discover the communities of collaborators. He founds that there are four distinct groups of collaborators dealing with the topic at hand.

\subsubsection{Selecting reviewers} 

Once Robert identifies one or more candidate reviewers who fit his requirements, he inputs their names in the \emph{Selected Reviewers} field (also with the help of the drop-down menu). He first decides to chose Pierre Paulin, a senior researcher. The colouring of the selected reviewer switches to blue both in the Researcher Timeline and the Researcher Network, and the colouring of his co-authors switches to grey, to identify them as conflicting potential reviewers, and tell them apart from the remaining available candidates. Then, Robert evaluates Hujun Bao, whose expertise fits with his requirements, then he decides to go for a younger researcher, and selects one of Bao's younger collaborators, Jin Huang. Among the remaining candidates, Robert chooses Xiao-Ming Fu, because he belongs to a different community than the previous two, and he has been working very recently on the subject at hand. 

Robert downloads his list of three reviewers with a click on the download button. The list reports reviewers' names and bibliographic references to their papers. 

After contacting the reviewers, Robert finds that one of them declines his invitation. Fortunately, for each reviewer selected by Robert, ReviewerNet has automatically added a list of potential alternative reviewers, in case of a negative answer by the original reviewer. Alternative reviewers are chosen from the candidate ones, so that they only conflict with the declining reviewer. Robert evaluates possible substitutes, again taking advantage of ReviewerNet functionalities, and finds his best replacement.  

\subsubsection{Discussion} 

This abreviated scenario shows how ReviewerNet can support investigating the literature, learning who are the experts in a field, and exploring relationships among them. The description above necessarily simplified a typical intercation process: Robert could of course switch back and forth between different tasks. For example, he could have refined the Paper Network after having examined the list of candidate reviewers. He could have adjusted the size of the list by fine tuning the parameters defining the criteria on productivity to be included in the list, or the criteria that defined conflicts. The process is iterative in nature, and the desiderata may evolve as the search proceeds. Thanks to the user-friendly interface which leaves the user control over the process, ReviewerNet enables the user to narrow down as well as widen the scope of analysis. In turn, the combined visualization of different aspects of the problem at hand well supports the decision making process.

\section{Evaluation}\label{sec:evaluation}

\begin{table}[!t]
	\renewcommand{\arraystretch}{1.3}
	\caption{Information about the 15 participants in the user study.}
	\label{table:infotesters}
	\begin{subtable}[t]{.99\linewidth}
		\centering%
		\begin{tabular}{|c|c|c|c|}
			\hline
			\ & PhD student & $\leq$ 12 & $>$ 12\\
			\hline
			{\bf Years from PhD} & 0.0\% & 66.7\% &  33.3\% \\
			\hline
		\end{tabular}
  \end{subtable}
	\par\bigskip
  \begin{subtable}[t]{.99\linewidth}
		\centering
		\begin{tabular}{|c|c|c|c|}
			\hline
			\ & $<$ 10 & 10 - 20 & $>$ 20\\
			\hline
			{\bf Reviews per year} & 6.7\% & 40.0\% &  53.3\% \\
			\hline
		\end{tabular}
  \end{subtable}
	\par\bigskip
  \begin{subtable}[t]{.99\linewidth}
		\centering
		\begin{tabular}{|c|c|c|}
			\hline
			\ & $\leq$ 3 & $>$ 3\\
			\hline
			{\bf Reviewer selections in 2018} & 13.3\% &  86.7\% \\
			\hline
		\end{tabular}
  \end{subtable}
\end{table}



We evaluated ReviewerNet on a dataset, described in Section \ref{sec:demonstration}, focused on Computer Graphics. We decided to ask the scientific community directly, that is, to involve real end-users, instead of in-house testers. We sent an email to the 60 members of the IPC of Eurographics Conference 2018, and to additional experts with a record of publications in the top venues of the sector. None of the subjects were involved in the work on ReviewerNet, and none of them knew the system prior to the evaluation test. The participation was on a volunteer basis, with no reward. 

We collected 7 responses from the IPC members (10\% of the IPC) and 8 responses from additional experts for a total of 15 users. The questionnaire was anonymous and the volunteers were asked to answer three questions about themselves: number of years from their PhD, reviews and reviewer selections per year; Table \ref{table:infotesters} shows the distribution of the results of this part of the questionnaire. 

The volunteers were asked to search three reviewers for a paper that they had to choose reviewers for in the recent past. This was done so that we could not only collect feedback on the system itself, but also enable the volunteers to comparatively evaluate the performance of the system.

For training, the volunteers were only provided with a 6-minutes video demonstrating the usage of ReviewerNet, namely the video recording the scenario in Section \ref{sec:demonstration}. We did not give any additional training. Also, we asked for a response within five days. This was done to evaluate whether it was easy to get acquainted with ReviewerNet, and whether the system was intuitive and quick to learn. Only one user out of 15 (6.7\% of the sample) reported s/he was not able to figure out how to use the system. The other 14 (93.3\%) were able to complete the task assigned with the little support offered. This confirms the user-friendliness of the instrument even if the tool offers many different interaction modalities.

The rest of the questionnaire was divided in two sections, whose questions and summary of answers are reported in Table \ref{table:formsection1} and Table \ref{table:formsection2}, respectively. The first section asked the user's opinion about the different functionalities of ReviewerNet, namely: finding key papers (and hence key researchers); presenting the scientific career of candidate reviewers; avoiding conflicts of interest; and finding sets of well distributed reviewers:
\begin{itemize}
\item [73.3\%] of the testers evaluated ReviewerNet as either good or excellent in finding key papers and researchers. One of the testers observed that {\em "[...] inserting manually key papers, takes a little more time, but then the system helps a lot navigating trough related papers and authors"};
\item [80.0\%] of the testers evaluated as good or excellent the presentation of the scientific career of candidate reviewers. One of the testers found that {\em "[...] the timeline also is a great added value with respect to imagining whether an author is doing a similar research now or he did many years ago"};
\item [ 86.7\%] of the testers thought ReviewerNet was good or excellent to help avoiding conflicts of interest; 
\item [ 66.7\%] evaluated as good or excellent ReviewerNet support to find sets of well distributed reviewers. One of the testers found  {\em "[...] a little difficult the interpretation of the researchers network  
[...] but probably again it is just a matter of more practice}".
\end{itemize}

The second section of the questionnaire asked the users an overall opinion on ReviewerNet, in terms of improvement of the overall quality of the reviewing process, and reduction in the time spent to search for reviewers:  
%
\begin{itemize}
\item [71.4\%] of the testers agreed or strongly agreed that ReviewerNet helps choosing good sets of reviewers, and hence improves the overall quality of the reviewing process;
\item [71.4\%] agreed or strongly agreed that ReviewerNet reduces the time spent to look for good sets of reviewers. 
\end{itemize}

In addition, the users could insert additional comments about ReviewerNet strengths and weaknesses, and suggestions for improvement.
One of the testers observed how {\em "[...] the tool actually follows my current practice, that is, look among authors of key papers"} but with the added value of the explicit labeling of conflicting reviewers. He/she also observed that {\em "[...] the labeling of conflicting reviewers helps also a lot. [...] the tool also helps in selecting reviewers from different areas, covering better the topic of a paper."}


\begin{table*}[!t]
	\renewcommand{\arraystretch}{1.3}
	\caption{ Distribution of answers to the first section of the questionnaire (14 participants).}
	\label{table:formsection1}
    \centering%
		\begin{tabular}{|l|c|c|c|c|c|}
			\hline
			& Very poor & Poor & Average & Good & Excellent\\
			\hline 
			{{\em How do you rate ReviewerNet in finding key papers and researchers?}}                      & 0.0\% & 0.0\% &  26.7\% & {\bf 46.6\%} & 26.7\% \\
			\hline
			{{\em How do you rate ReviewerNet in presenting the scientific career of candidate reviewers?}} & 0.0\% & 6.7\% &  13.3\% & {\bf 46.6\%} & 33.3\% \\
			\hline
			{{\em How do you rate ReviewerNet in avoiding conflicts of interest?}}                          & 0.0\% & 6.7\% &   6.7\% & 40.0\% & {\bf 46.7\%} \\
			\hline
			{{\em How do you rate ReviewerNet in finding sets of well distributed reviewers?}}              & 0.0\% & 6.7\% &  26.7\% & {\bf 40.0\%} & 26.7\% \\
			\hline
		\end{tabular}
\end{table*}

\begin{table*}[!t]
	\renewcommand{\arraystretch}{1.3}
	\caption{Distribution of answers to the second section of the questionnaire (13 participants).}
	\label{table:formsection2}
    \centering%
		\begin{tabular}{|p{8cm}|c|c|c|c|c|}
			\hline
			& Strongly disagree & Disagree & Neutral & Agree & Strongly agree\\
			\hline 
			{{\em I think that ReviewerNet helps choosing good sets of reviewers, and hence improves the overall quality of the reviewing process}.} & 0.0\% & 0.0\% &  28.6\% & {\bf 35.7\%} & {\bf 35.7\%} \\
			\hline
			{{\em I think that ReviewerNet reduces the time spent to look for good sets of reviewers}.}                                              & 0.0\% & 0.0\% &  28.6\% & 28.6 & {\bf 42.9} \\
			\hline 
		\end{tabular}
	\par\bigskip
\end{table*}

Concerning ReviewerNet ability to find key papers and researchers, one of the testers observed that vision-related venues (e.g., conferences such as CVPR and ICCV and journal such as IJCV and TPAMI) were missing from the list of sources for key papers and authors on which the demonstration tool was built. He/she observed that the tool would have been more useful if these were included, since many works overlap vision and graphics. Similarly, another tester observed that the homogeneous nature of the sources selected made so the proposed reviewers could show no enough divergence and could be scarce. In this respect, it is worth noticing that ReviewerNet can be built over \emph{any} subset of the Semantic Scholar corpus and customized to include the venues of interest. Therefore, these comments mostly apply to the particular instance used for testing.

In the light of these considerations, we believe the results from the evaluation study show a very high value of user satisfaction, and also potential room for improvement.


\section{Conclusions}
\label{sec:conclusions}

We have presented ReviewerNet, a novel system for choosing reviewers by visually exploring scholarly data. ReviewerNet enables scientific journal editors and members of IPCs to search the literature about the topic of a submitted paper, to identify experts in the field and evaluate their stage of career, to check possible connections with the submitting authors and among the reviewers themselves, helping therefore to avoid conflicts and to build a fairly distributed pool of reviewers. To do so, ReviewerNet features a combined visualization of the literature, the career of potential reviewers, their conflict of interests, and their nets of collaborators. 

The results from a user study involving 15 senior members from the Computer Graphics community confirmed that they were able to get acquainted with the system even with a very limited training, and  appreciated the different functionalities of ReviewerNet and its capability of improving the reviewer search process.

Interestingly enough, the system is able to help the process even without exploiting any content-based analysis of the papers. While it is true that there is room for improving the system by partially automating the choice of the key papers, in its current state ReviewerNet focuses on the deterministically mechanical part of the process, minimizing the possibility of introducing any bias in the process. 

In the future, we plan to discuss with providers of scholarly data about the possible release of a version of ReviewerNet with customizable data coverage, to be used by the various scientific communities.

\ifCLASSOPTIONcompsoc
  \section*{Acknowledgments}
\else
  \section*{Acknowledgment}
\fi

The authors would like to thanks the Allen Institute for Artificial Intelligence for providing the Semantic Scholar corpus of bibliographic data.

\ifCLASSOPTIONcaptionsoff
  \newpage
\fi



\bibliographystyle{IEEEtran}
\bibliography{ReviewerNetBib}
%



%

\begin{IEEEbiographynophoto}{Mario Salinas}
Mario Leonardo Salinas has earned the bachelor degree in Computer Science in 2017 at University of Pisa where he is currently studying for the Computer Science and Data Science Master Degree. His main academic interests and projects are about Big Data Analytics and Data Visualization.
\end{IEEEbiographynophoto}

\begin{IEEEbiographynophoto}{Daniela Giorgi}
Daniela Giorgi is a Researcher at CNR-ISTI. She received a Degree cum Laude in Mathematics from the University of Bologna in 2002, and a Ph.D. in Applied Mathematics from the University of Padua in 2006. Her research interests include computational topology and geometry for shape analysis and comparison. She has authored about 60 peer-reviewed publications in journals and conferences, and a book about mathematical tools for 3D data analysis.
\end{IEEEbiographynophoto}


\begin{IEEEbiographynophoto}{Paolo Cignoni}
Dr. Paolo Cignoni is a Research Director with CNR-ISTI. He received a Ph.D. Degree in Computer Science at the University of Pisa in 1998. He has been awarded "Best Young Researcher" by Eurographics in 2004 and he is a Fellows of the Eurographics Association since 2016. His research interests cover many Computer Graphics fields as geometry processing, 3D scanning data processing, digital fabrication, scientific visualisation and digital heritage. He has published more than 150 papers in international refereed journals/conferences, has an h-index of 47, and has served in the Program committee of all the most important conferences of Computer Graphics. 
\end{IEEEbiographynophoto}




\end{document}